\documentclass[aps,pra,twocolumn,showpacs,amsfonts,amssymb,amsmath]{revtex4}

\usepackage{graphicx}
\usepackage{dcolumn}

\begin{document}
\preprint{draft 1.0}

\title{Observation of Electric Quadrupole Transitions to Rydberg \textit{nd} States of Ultracold Rubidium Atoms}

\author{D. Tong}
\author{S. M. Farooqi}
\author{E. G. M. van Kempen}
\author{Z. Pavlovic}
\author{J. Stanojevic}
\author{R. C\^ot\'e}
\author{E. E. Eyler}
\author{P. L. Gould}
\affiliation{Physics Department, University of Connecticut, Storrs, CT}

\date{\today}

\begin{abstract}
We report the observation of dipole-forbidden, but quadrupole-allowed, one-photon transitions to high Rydberg states in Rb. Using pulsed UV excitation of ultracold atoms in a magneto-optical trap, we excite $5s \rightarrow  nd$ transitions over a range of principal quantum numbers $n=27-59$. Compared to dipole-allowed (E1) transitions from $5s \rightarrow np$, these E2 transitions are weaker by a factor of approximately 2000. We also report measurements of the anomalous $np_{3/2} : np_{1/2}$ fine-structure transition strength ratio for $n=28-75$. Both results are in agreement with theoretical predictions.
\end{abstract}

 \maketitle


\section{Introduction}

Samples of ultracold atoms have a number of benefits for spectroscopic measurements. The low velocities result in reduced Doppler shifts and long interaction times, significantly reducing both Doppler broadening and transit-time broadening. In addition, the atoms can be highly localized and prepared in specific states. Microwave \cite{Li03} and optical transitions \cite{Khaykovich00, Marian05, Fortier06} involving excited atomic states have been investigated with ultracold samples, as have photoassociative processes \cite{Jones06}, which probe bound molecular states. An example closely related to the results presented here is the observation of the Na $3p \rightarrow 4p$ quadrupole transition in an ultracold sample \cite{Bhattacharya03}.

In the present work, we use pulsed UV excitation of ultracold Rb atoms to measure the oscillator strengths of weak single-photon transitions from the $5s$ ground state to $nd$ Rydberg states, where $n \gg 1$ is the principal quantum number. In zero electric field, there is no Stark mixing and these $5s \rightarrow nd$ transitions are dipole (E1) forbidden, but quadrupole (E2) allowed. Such transitions must be considered when using excitation spectra to probe external electric fields or the interactions between ultracold Rydberg atoms. Also, these E2 transitions are potentially useful in extending the number of Rydberg states that can be excited. For example, two-frequency UV light could be used to simultaneously excite Rb $5s$ atoms to $np$ and $(n-1)d$ states via one-photon transitions. Pairs of atoms in these states will have strong dipole-dipole interactions \cite{Afrousheh04,Li05}. In previous work with Rb, E2 oscillator strengths to $nd$ states for $n=4-9$ \cite{Niemax77} and for $n=4$ \cite{Nilsen78} were measured. By contrast, we probe states of much higher $n$: $n=27-59$. We determine absolute E2 oscillator strengths by combining our measured ratios of signals for the $5s \rightarrow nd$ (E2) and $5s \rightarrow (n+1)p$ (E1) transitions with previously determined absolute E1 oscillator strengths \cite{Shabanova84}. To compare with our measurements, we calculate the E2 oscillator strengths for high $n$ using phase-shifted Coulomb wavefunctions. This comparison yields reasonably good agreement.

We also present measurements of the anomalous ratio of E1 transition strengths to the two $np$ fine-structure states, $J$=3/2 and $J$=1/2. This ratio is a sensitive probe of electronic wavefunctions in many-electron atoms. Our results, covering the range $n=28-75$, are found to be consistent with a number of previous calculations and with previous measurements, most of which involve lower $n$.

In Sect. II, the E2 oscillator strength calculations are described. In Sect. III, we describe the experimental setup. Measurements of E2 transition strengths and E1 fine-structure transition strength ratios are described and compared with theory in Sects. IV and V, respectively. Sect. VI comprises concluding remarks.

\section{Theory}

We first compute the E2 excitation probability for a $5s\rightarrow nd$
transition by a linearly polarized laser pulse of finite duration.
We start with the Hamiltonian for the electric quadrupole interaction \cite{Bethe77}
\begin{equation}
     H_Q = \frac{e}{2} \left( \vec{E}_d\cdot\vec{r}\right)
             \left( \vec{k}\cdot\vec{r}\right) \; ,
\label{eq:H_Q}
\end{equation}
where $\vec{E}_d$ is the electric field, $\vec{r}$ the
position of the electron, and $\vec{k}$ the wave number
of the electromagnetic field.

In the present experiment the ground-state atoms have no alignment or orientation,
so the excitation probability is not polarization-dependent.
This allows us to neglect the hyperfine structure and to start with a $5s_{1/2}$
initial state, which is not capable of alignment.  We nevertheless treat in explicit
form the case of linear polarization, so that our results can readily be generalized
to more complicated cases.

For simplicity, we choose the
electric field along $z$, and the wave travelling
along $x$, {\it i.e.} $\vec{E}_d = E_d(t) \hat{e}_z$
and $\vec{k} = k \hat{e}_x$, where $E_d(t)=E_{d0} \cos \omega t$.
Within the rotating-wave approximation, Eq.(\ref{eq:H_Q})
becomes
\begin{equation}
    H_Q = \frac{e}{2}E_d(t)k xz = \frac{eE_{d0}k}{4} xz \;,
\label{eq:H_Q-RWA}
\end{equation}
and using $xz = -r^2 \sqrt{\frac{2\pi}{15}} \left( Y_2^1 - Y_2^{-1}
\right)$, it can be written in term of spherical harmonics $Y_\ell^m$ as
\begin{equation}
    H_Q = \frac{eE_{d0}k}{4} r^2 \sqrt{\frac{2\pi}{15}}
              \sum_{m=-\frac{1}{2},\frac{3}{2}} a_m Y_2^{m-\frac{1}{2}}  \; ,
\label{eq:H_Q-Ylm}
\end{equation}
with $a_{m=-\frac{1}{2}}=+1$ and $a_{m=+\frac{3}{2}}=-1$.

For Rb in a MOT, the ground state
$|5\;s_{1/2}\;m_j= \pm \frac{1}{2}\rangle$ can only be excited to
$d$ states via E2 transitions. We consider here the case $m_j= \frac{1}{2}$;
 identical results follow for $m_j= -\frac{1}{2}$. From the selection rules,
  $m_j$ changes by $\pm 1$, and the allowed $d$ states (with $j=\frac{5}{2}$ or
$\frac{3}{2}$) are $|n\; d_j\; m_j=-\frac{1}{2}\rangle$ and $|n\; d_j\;
m_j=\frac{3}{2}\rangle$. If we label the amplitude in the ground $5s$
state by $c_s$ and the amplitude in the excited states $|n\;d\;m\rangle$
with $m=-\frac{1}{2}$ or $\frac{3}{2}$ by $c_m$,
we get the following Bloch equations
\begin{eqnarray}
i\frac{dc_s}{dt} & = & \sqrt{\frac{2\pi}{15}} \frac{eE_{d0} k}{4\hbar}
                        \langle 5s||r^2||nd\rangle \nonumber \\ & & \times
                        \sum_{m=-\frac{1}{2},\frac{3}{2}} \left\langle \frac{1}{2},\frac{1}{2}\left|
                               a_{\frac{1}{2}-m}Y_2^{\frac{1}{2}-m}
                               \right|jm\right\rangle c_m\\
i\frac{dc_m}{dt} & = & \sqrt{\frac{2\pi}{15}} \frac{eE_{d0} k}{4\hbar}
                        \langle nd||r^2||5s\rangle \nonumber \\ & & \times
                        \left\langle jm \left|a_{m-\frac{1}{2}} Y_2^{m-\frac{1}{2}}
\right|\frac{1}{2},\frac{1}{2}\right\rangle c_s
\end{eqnarray}
If we define
\begin{equation}
   c_{d_j} \equiv \sum_{m=-\frac{1}{2}, \frac{3}{2}}
                  \left\langle \frac{1}{2},\frac{1}{2}\left|
                               Y_2^{\frac{1}{2}-m}
                               \right|jm\right\rangle a_{\frac{1}{2}-m}c_m \; ,
\end{equation}
and
\begin{eqnarray}
  W & \equiv & \sqrt{\frac{2\pi}{15}} \frac{eE_{d0} k}{2\hbar}
                        \langle nd||r^2||5s\rangle \;, \\
  \beta^2_j & \equiv & \sum_{m=-\frac{1}{2},
\frac{3}{2}}a_{m-\frac{1}{2}}a_{\frac{1}{2}-m}
                     \left\langle
\frac{1}{2},\frac{1}{2}\left|Y_2^{\frac{1}{2}-m}
                   \right|jm \right\rangle \nonumber \\ & &
\times
                   \left\langle jm \left|Y_2^{m-\frac{1}{2}}
                   \right|\frac{1}{2},\frac{1}{2}\right\rangle \; ,
\end{eqnarray}
we can rewrite the Bloch equations as
\begin{eqnarray}
  i\frac{dc_s}{dt} & = & \frac{W}{2} c_{d_j} \;, \\
  i\frac{dc_{d_j}}{dt} & = & \beta_j^2 \frac{W}{2} c_s \;.
\end{eqnarray}
The solutions are simply
\begin{eqnarray}
   c_s(t) & = & \cos \frac{W\beta_j t}{2} \;, \\
   c_{d_j}(t) & = & -i\beta_j\sin \frac{W\beta_j t}{2} \;,
\end{eqnarray}
and the quadrupole excitation probability $P^Q_j = 1 - |c_s|^2$ has the form of a Rabi equation,
\begin{equation}
  P^Q_j = \sin^2 \frac{W\beta_j t}{2}\simeq \frac{W^2\beta_j^2 t^2}{4}\; .
\end{equation}
The result on the right assumes a short interaction time.
We also note that $W$ is assumed to be real here ({\it i.e.} no chirp).

After some algebra, and using $a_{m-\frac{1}{2}}a_{\frac{1}{2}-m}=-1$,
the expression for $\beta_j^2$ for $j=\frac{5}{2}$ and $\frac{3}{2}$ is
found to be
\begin{equation}
   \beta^2_j = \frac{2j+1}{20\pi} \;.
\end{equation}
Therefore, ignoring the $j$-dependence of $\langle nd||r^2||5s\rangle$,
we recover the statistical ratio of the 5/2 and 3/2 components for E2
excitations
\begin{equation}
   \frac{P_{j=\frac{5}{2}}^Q}{P_{j=\frac{3}{2}}^Q}
   \simeq  \frac{W^2\beta^2_{\frac{5}{2}} t^2/4}{
W^2\beta^2_{\frac{3}{2}}t^2/4}
   = \frac{3}{2} \;.
\end{equation}

To compare the measured E2 signal for a $5s\rightarrow n_dd$ quadrupole
transition with a nearby E1 $5s\rightarrow n_pp$ dipole transition, we
compute the ratio $\eta$ between the E2 and E1 excitation probabilities,
assuming
$P_{j_p}^{\rm dip}=\sin^2 \omega_{j_p} t/2$ and the same pulse duration
$t$ for both cases:
\begin{equation}
  \eta_{j_p,j_d} = \frac{P_{j_d}^Q}{P_{j_p}^{\rm dip}}
       \simeq \frac{W^2\beta^2_{j_d}}{\omega^2_{j_p}}
\end{equation}
For $5s\rightarrow n_p p_j$ transitions with polarized light
$(\vec{E}_p=E_p\hat{e}_z)$, we have for the E1 Rabi frequency,
\begin{equation}
   \omega_{j_p} = \frac{eD_{j_p} E_{p0}}{\hbar}
\end{equation}
with
\begin{equation}
  e D_{j_p} = \sqrt{f_{j_p} \frac{6\hbar e^2}{2 m_e 2\pi\nu}
              \left(\begin{array}{ccc} \frac{1}{2} & 1 & j_p \vspace{2 mm} \\
                    \frac{1}{2} & 0 & -\frac{1}{2} \end{array}\right)^2}
\end{equation}
where $f_{j_p}$ is the oscillator strength to the $j_p$ component,
and $\nu$ is the transition frequency in Hz.
 From the definition of $W$ and the result for $\beta^2$, we
have

\begin{eqnarray}
  \eta_{j_p,j_d} & = &
\frac{(2j_d+1)}{150}\frac{E_{d0}^2}{E_{p0}^2}\frac{k^2}{4D_{j_p}^2}
               |\langle nd||r^2||5s\rangle|^2 \;, \\
       & = & 6.421\times 10^{-10} \frac{(2j_d+1)}{f_{j_p}} \frac{I_d}{I_p}
             |\langle nd||r^2||5s\rangle|^2
       \label{eq:eta_j}
\end{eqnarray}
where $\langle nd||r^2||5s\rangle$ is in atomic units, and $I_d$ and $I_p$
are the laser intensities used to excite the $n_pp$ and $n_dd$ states,
respectively.  We also use the fact that for $k = 2\pi\nu/c$, and for these high Rydberg states
we take $h \nu \simeq h(1.010 \times 10^{15}$ Hz) as equal to the $5s$ ionization energy.

It is easy to modify Eq. \ref{eq:eta_j} to find the ratio $\eta$ of the total signal sizes summed over fine-structure components.  This requires only summation over $j_d = \frac{5}{2}$ and $\frac{3}{2}$ in the numerator, and replacement of $f_{j_p}$ in the denominator by the total E1 oscillator strength to the $p$ state, $f_p$:
\begin{equation}
  \eta =  \frac{6.421\times 10^{-9}}{f_p} \frac{I_d}{I_p}
             |\langle nd||r^2||5s\rangle|^2
       \label{eq:eta}
\end{equation}

Finally it is useful to relate the ratio $\eta$, which in general is polarization-dependent, to the ratio of oscillator strengths $f_d(n_d)/f_p(n_p)$, which is not.  In the present experiment the absence of alignment in the initial state nullifies this distinction, and it is easily shown that $\eta = f_d(n_d)/f_p(n_p)$, where the E2 oscillator strength is given by $f_d(n) = 6.231 \times 10^{-54} \nu^3 \langle nd||r^2||5s\rangle|^2$.

\begin{table}
\caption{Electric quadrupole matrix elements and oscillator strengths $f_d$ for $5s \rightarrow nd$ transitions.  Matrix elements are in units of $a_0^2$ and oscillator strengths are in units of $10^{-10}$. Columns 2 and 3 show calculated values.  Experimental values in column 4 are obtained from measured signal size ratios $\eta$ as described in Section IV.}

\begin{ruledtabular}
\begin{tabular}{c c c l}

 $n$  & $|\langle nd||r^2||5s \rangle|^2$ & $f_{d,\rm{calc}} (10^{-10})$ & $f_{d,\rm{expt}} (10^{-10})$ \\
\hline
 24  &  $1.14 \times 10^{-1}$ & 7.15 &  \\
 27  &  $7.84 \times 10^{-2}$ & 4.96 & $7.45 \pm 2.0 $ \\
 29  &  $6.27 \times 10^{-2}$ & 3.97 & $6.15 \pm 1.9 $ \\
 34  &  $3.82 \times 10^{-2}$ & 2.43 & $4.29 \pm 1.6 $ \\
 39  &  $2.49 \times 10^{-2}$ & 1.59 & $2.68 \pm 0.44 $ \\
 44  &  $1.72 \times 10^{-2}$ & 1.10 & $1.84 \pm 0.28 $ \\
 47  &   &  & $1.67 \pm 0.27$ \\
 49  &  $1.23 \times 10^{-2}$ & 0.79 & $1.27 \pm 0.33$ \\
 54  &  $9.13 \times 10^{-3}$ & 0.58 & $1.38 \pm 0.33$ \\
 59  &  $6.95 \times 10^{-3}$ & 0.45 & $1.33 \pm 0.22$ \\

\end{tabular}
\end{ruledtabular}
\end{table}

Although the E1 oscillator strengths to the high-$n$ $p$ states are well-known \cite{Shabanova84}, the quadrupole oscillator strengths $f_d$ are not.  To evaluate them numerically for comparison with our experiment, we have computed the quadrupole matrix element $|\langle nd||r^2||5s\rangle|^2$
using the model potential for Rb $5s$ produced by Marinescu {\it et al.} \cite{Marinescu94} and phase-shifted Coulomb wavefunctions for the Rydberg states.
This precise one-electron model potential was developed to represent the
motion of the valence electron in the field of the closed alkali-metal
positive-ion core.  In Table I we show the calculated matrix elements and the corresponding values of the total oscillator strength $f_d$.  To evaluate their accuracy, we compare them with measured values
of E2 transitions by Niemax \cite{Niemax77}
and calculated values by Warner \cite{Warner68} for low $n_d$ states.
For $n_d<10$, our results differ by 20-35\% from those
of Warner, and by 40-70\% with those of Niemax, although
the agreement with Niemax's values seems to get
better rapidly with growing $n_d$, falling to 5\% for
$n_d=9$. Overall, we estimate the error of our calculated
oscillator strengths to be roughly 40-60\%.

\section{Experiment}
In our experiments, we use pulsed UV excitation to probe high Rydberg states of ultracold $^{85}$Rb atoms. These pulses are generated by pulsed amplification of cw light at $\sim$594 nm followed by frequency doubling in a BBO crystal. The cw light is generated by a single-frequency tunable ring dye laser system (Coherent 699-29 with Rhodamine 6G dye) pumped by an argon-ion laser. This cw light seeds a three-cell amplifier chain in which the capillary dye cells are pumped by the second harmonic (532 nm) of an injection-seeded pulsed Nd:YAG laser. The frequency-doubled pulses at $\sim$297 nm are typically 5 ns in duration (FWHM) and have energies up to 500 $\mu$J/pulse. The UV bandwidth of $\sim$140 MHz, measured by scanning over the $30p$ resonance at low UV intensity (18 kW/cm$^2$), is roughly twice the Fourier transform limit, and is determined by optical phase perturbations in the pulsed amplifier \cite{Melikechi94}.

The ultracold (T $\sim$100 $\mu$K) sample of $5-10 \times 10^6$ Rb atoms is generated in a diode-laser-based vapor-cell magneto-optical trap (MOT) in which densities up to 10$^{11}$ cm$^{-3}$ are achieved. Rydberg excitation is performed by focusing the UV light into the MOT cloud, yielding a cylindrical excitation volume $\sim$500 $\mu$m long and $\sim$220 $\mu$m in diameter (FWHM). To prevent direct photoionization from the $5p_{3/2}$ state by the UV light, the trapping and repumping beams are turned off with acousto-optical modulators about 2 $\mu$s before the UV pulse arrives. Usually, the trapping light is switched off slightly ($\sim$500 ns) before the repumping light in order to ensure that all atoms are in the $5S_{1/2}(F=3)$ level when the UV pulse arrives. However, we can vary the effective atomic density without affecting other properties of the MOT, by delaying the turn-off of the repump laser \cite{Tong04}. This allows the trapping light to optically pump atoms from $F$=3, the hyperfine level probed by the UV laser, into $F$=2. The time scale for this population transfer is $\sim$100 $\mu$s.

The MOT is located between a parallel pair of 95\% transparent grids separated by 2.09 cm. These grids allow control of the static electric field during Rydberg excitation as well as pulsed-field ionization of the Rydberg atoms and extraction of the resulting ions. The applied field is perpendicular to the linear polarization of the UV light. Within 100 ns after the Rydberg atoms are created, a pulsed field of $\sim$1500 V/cm is applied, ionizing states with principal quantum numbers as low as $n$=25. The ions are detected with a discrete dynode electron multiplier (ETP model 14150). A boxcar averager is used to select the desired time-of-flight window.

\begin{figure}
\centering \vskip 0 mm
\includegraphics[width=0.95\linewidth]{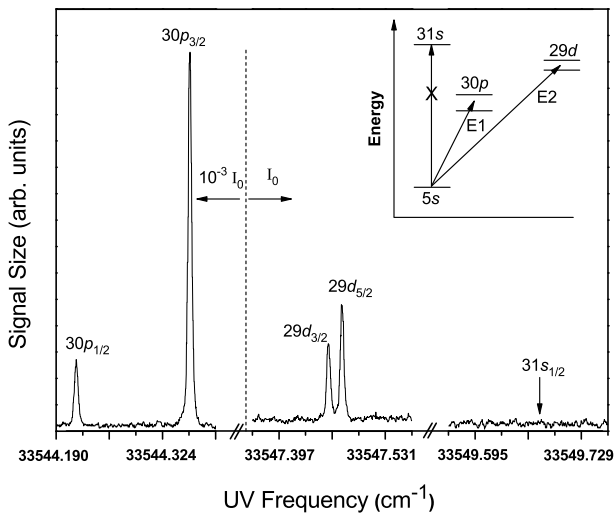}
\caption{\protect\label{fig1} Spectrum showing $5s_{1/2} \rightarrow 30p_J$ (E1) and $5s_{1/2} \rightarrow 29d_J$ (E2) transitions with no external electric field. The location of the $5s_{1/2} \rightarrow 31s_{1/2}$ transition, which is both E1 and E2 forbidden, is indicated by the arrow. The laser intensity for the 29$d$ and 31$s$ portions of the scan is $I_0$=18.8 MW/cm$^2$, while for the 30$p$ region, the intensity is reduced by a factor of 1000 to avoid saturation effects. The inset shows the energy levels (not to scale) and associated E1 and E2 transitions.}
\end{figure}

A typical excitation spectrum for $n$=30 is shown in Fig. 1. The $5s_{1/2} \rightarrow np_J$ E1 transitions are visible, as well as the nearby $5s_{1/2} \rightarrow (n-1)d_J$ E2 transitions. Transitions to $(n+1)s_{1/2}$ states are not observed because they are both dipole- and quadrupole-forbidden. For this scan, the stray electric fields were minimized as described below, resulting in negligible Stark mixing. Thus the entire $5s_{1/2} \rightarrow nd_J$ signal can be attributed to the E2 transition.

At a given value of $n$, the E1 and E2 transitions we measure have oscillator strengths that differ by three orders of magnitude. In addition, their strengths vary significantly with $n$. To avoid possible problems with either the limited dynamic range of our detector, or saturation of the transitions themselves, we adjust the UV intensity to maintain similar signal sizes for all scans. Low intensities (e.g., 140 kW/cm$^2$ for $n$=28) are used to observe the relatively strong E1 transitions ($5s_{1/2} \rightarrow np_J$), while higher intensities (e.g., $\times$1000) are used to drive the weaker E2 transitions to $(n-1)d_J$. At each $n$, the ratio of oscillator strengths is obtained by dividing the measured E2:E1 signal ratio by the ratio of intensities used. A polarizer and half-wave plate combination and neutral density filters are used to vary the intensity. The relative intensities used for each scan are directly measured to within 5\% with a UV photodiode.  Using this scheme, the residual errors due to detector nonlinearity for comparing the signal strengths of the various fine-structure components are no more than 2-3\%.

\begin{figure}
\centering \vskip 0 mm
\includegraphics[width=0.95\linewidth]{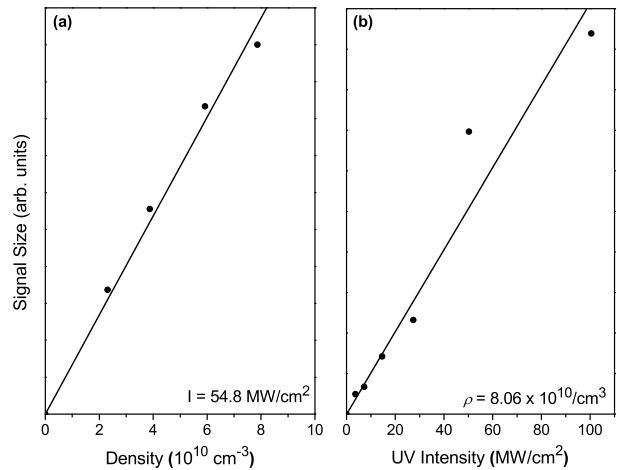}
\caption{\protect\label{fig2} (a) 59$d$ signal as a function of atomic density at a fixed intensity of 55 MW/cm$^2$. (b) 59$d$ signal as a function of intensity at a fixed density of $8.1 \times 10^{10}$ cm$^{-3}$. Linear fits through the origin are shown for both cases.}
\end{figure}

Since E2 transitions are one-photon processes occurring in individual atoms, the corresponding signals should be linear with respect to both atomic density and UV intensity. These dependencies are verified in Figs. 2(a) and 2(b), respectively. This verification is particularly important because it rules out the possibility that these $5s_{1/2} \rightarrow nd_J$ features are variations of previously observed two-photon molecular resonances \cite{Farooqi03,Stanojevic06}. The molecular resonances are due to avoided crossings between long-range molecular potentials of two Rydberg atoms and occur at the average energy of two states that are strongly coupled to $np$ states. For example, the $(n-1)d$ and $ns$ states are dipole coupled to $np$, leading to a molecular resonance at the average energy of these states. Since two excited atoms are involved, the molecular resonance signal is quadratic in both atomic density and UV intensity \cite{Farooqi03}. In principle, a molecular resonance of this type could occur at the frequency of the $5s_{1/2} \rightarrow nd_J$ transition via two-photon excitation of a pair of atoms to an $nd_J + nd_J$ configuration. The measured density and intensity dependencies indicate that our $nd_J$ signals are instead due to single-atom E2 transitions.

Because the Stark effect can mix closely-spaced $p$ and $d$ states, a stray electric field can induce an E1 transition amplitude to an $nd_J$ state.  We have carefully studied the effect of an applied electric field on our $nd$ signals and conclude that for $n<49$, Stark mixing is negligible, and for $49\leq n \leq 59$, corrections for the stray electric field can be made. Because the Stark mixing increases rapidly with $n$, our $nd$ signals are dominated by the stray field for $n>69$. We use the measurements at high $n$ to correct the signals at lower $n$.

\begin{figure}
\centering \vskip 0 mm
\includegraphics[width=0.95\linewidth]{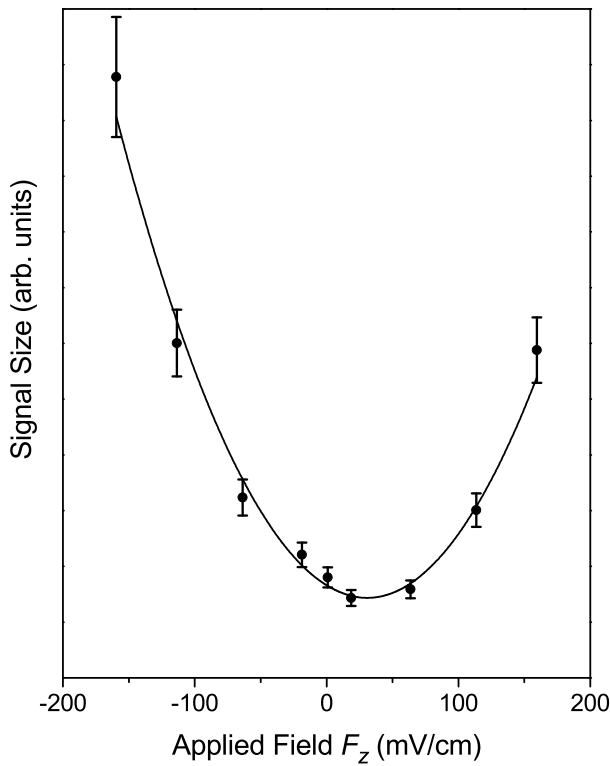}
\caption{\protect\label{fig3} 74$d$ signal as a function of applied electric field $F_z$. A least-squares fit to Eq. (1) is also shown. From the fit we find that the minimum signal occurs at $F_z$ = 31 mV/cm and that at this minimum, the residual stray field is 78 mV/cm.}
\end{figure}

Using the parallel grids in the MOT chamber, we are able to apply a uniform field $F_z$
and thus cancel any $z$-component of the stray field. However, we cannot
eliminate other components (or gradients) of the stray field. To determine the effects
of the residual stray field, we use high-$n$ states (e.g., $n$=74) to measure the $nd$ signal
as a function of $F_z$, as shown in Fig. 3. For small fields, the Stark-mixed $p$-state amplitude
is linear in $F_z$ and the measured $nd$ signal $S_d$ should be quadratric in $F_z$:
\begin{equation}
S_d = S_0 + \alpha(F_z+F_{z0})^2
\label{eq:StarkSignal}
\end{equation}
Fitting the data to the Stark parabola of Eq.~\ref{eq:StarkSignal} yields the $z$-component of the stray field $F_{z0}$, the Stark coefficient $\alpha$, and the minimum signal $S_0$. After having determined $F_{z0}$ = 31 mV/cm at several values of $n$ (69,74,84), we set $F_z$ to cancel this component of the stray field for subsequent scans at lower $n$.

The resulting minimum $nd$ signal, $S_0$ in Eq.~\ref{eq:StarkSignal}, can be written as $S_0 = S_{0F} + S_{\textrm{E2}}$, where $S_{0F}$ is the contribution from Stark mixing due to components of the stray field that are not canceled, and $S_{\textrm{E2}}$ is the E2 signal that we are trying to extract. These two contributions are dominant at high and low $n$, respectively. At a fixed UV intensity, the Stark-mixed $nd$ signal $S_{0F}$ scales as $(n^*)^7$ \cite{Gallagher94}, where the effective principal quantum number $n^* = n - \delta$  and the quantum defect $\delta$ is 1.3472 for high-$n$ $d$ states \cite{Lorenzen83}. Meanwhile, the $(n+1)p$ signal $S_p$ scales as $(n^*)^{-3}$ at high $n$ \cite{Shabanova84}. The normalized Stark-mixed $nd$ signal $S_{0F}/S_p$ should therefore scale as $(n^*)^{10}$. Because of this rapid $n$ scaling, $S_{0F}/S_p$ is the dominant contribution to $S_0/S_p$ for $n\geq 69$. We isolate this Stark contribution for these high $n$'s by subtracting the normalized E2 contribution $S_{E2}/S_p$ from the measured values of $S_0/S_p$. Here we assume that $S_{E2}/S_p$ is given by its average low-$n$ value of $6 \times 10^{-4}$ (see Fig. 4), which amounts to 17\% of $S_0/S_p$ at $n$=69. We now correct the low-$n$ ($n\leq 59$) data for Stark mixing by fitting the $S_{0F}/S_p$ values for high $n$ ($n$=69-89) to the expected $(n^*)^{10}$ scaling and then extrapolating this scaling to lower $n$. Typically at $n$=59 the Stark correction is 34\%, but it drops rapidly to 10\% at $n$=49. The E2 subtraction from the high-$n$ data has a minimal effect on the Stark corrections to the lower-$n$ data, e.g., it changes the $n=59$ Stark correction from 39\% to 34\%. We note that the non-cancelable stray field determined from the high-$n$ data is given by $(S_{0F}/\alpha)^{1/2}$ and is typically 80-100 mV/cm.

\section{E2 Oscillator Strengths}

\begin{figure}
\centering \vskip 0 mm
\includegraphics[width=0.95\linewidth]{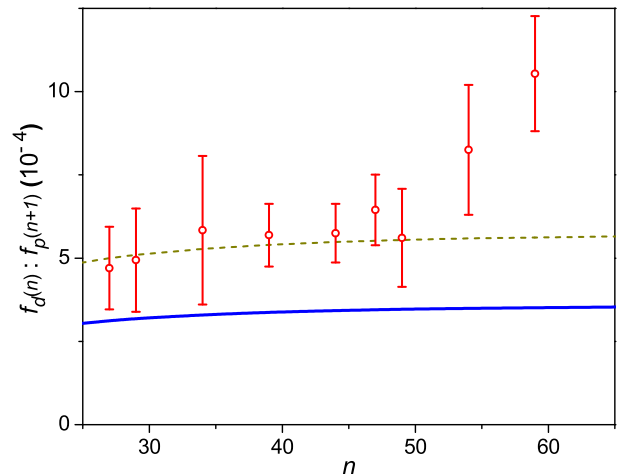}
\caption{\protect\label{fig4} (Color online) Ratio of E2 to E1 oscillator strengths as a function of $n$. The results for $n>49$ have been corrected for Stark mixing by the residual stray field. The solid line is the predicted ratio, and the dotted line is the prediction scaled by 1.6 for easier comparison with the data (see text).}
\end{figure}

In Fig. 4, we show the $n$ dependence of the ratio of the E2 oscillator strength $f_d(n)$ (for $5s \rightarrow nd$ transitions) to the E1 oscillator strength $f_p(n+1)$ (for the closest $np$ transition, $5s \rightarrow (n+1)p$). This ratio is obtained from the experimental data by taking the ratio of the $nd$ signal per unit intensity to the $(n+1)p$ signal per unit intensity. As discussed in Sect. III, we use higher (lower) intensities for the E2 (E1) transitions in order to have comparable signal sizes. The $nd$ and $(n+1)p$ signals are determined by summing the areas under each fine-structure peak using Gaussian fits, yielding oscillator strengths that are summed over $J$. The relative strengths for the different $J$'s are discussed in Sect. V. The results in Fig. 4 with $n>49$ have also been corrected for Stark mixing of the $nd$ states by the residual field as described in Sec. III. Uncertainties in the data include statistical contributions as well as contributions from signal area determinations, relative intensity measurements, and the Stark corrections. Over the range $n$=27-59, the ratio $f_d(n)/f_p(n+1)$ is seen to be relatively constant, with a possible slight increase with $n$.

Figure 4 also shows predicted oscillator strength ratios, obtained by combining the $f_d$ values in Table I with electric dipole oscillator strengths $f_p$ of Ref. \cite{Shabanova84}.  Using the extrapolation method shown in Fig. 2 of this paper, we find
\begin{equation}
f_p(n)  \simeq \frac{0.0234}{(n*)^3 } + \frac{1.58}{(n*)^5 },
\end{equation}
where $n^* = n-2.6415$ for the high-$n$ $p_{3/2}$ states \cite{Lorenzen83}.  These E1 oscillator strengths should be accurate to within at worst 15\%.  Although the calculated values are typically about 40\% smaller than the measurements, they are consistent given the 40-60\% uncertainty of the calculations.  Figure 4 also shows the calculated values scaled by a factor of 1.6, to aid comparison of the $n$-dependence between theory and experiment.  In this regard the agreement is good except at $n_d$=59, where the measured ratio is significantly larger than the calculated trend would indicated.  Because the Stark mixing correction is almost an order of magnitude larger at $n_d$=59 than at $n_d$=49, it is possible that this discrepancy arises because the Stark correction has been underestimated.

Because the E1 oscillator strengths are accurately known, it is also possible to obtain experimental determinations of the absolute E2 oscillator strengths $f_d$ from the data in Fig. 4.  The final column of Table I shows the resulting $f$ values. To the best of our knowledge, there are no previous high-$n$ $f_d$ measurements in Rb with which to directly compare our results. Niemax \cite{Niemax77} reports values for $n$=4-9, including $f_d = 1.3 \times 10^{-8}$ (summed over $J$) at $n$=9. Although we cannot extrapolate meaningfully to our range of high $n$, his $f$ values do decrease sharply with $n$. We note that the exceptionally large $f_d(n)/f_p(n+1)$ ratios we measure at high $n$ are due mainly to the anomalously small size of $f_p$, which results from the broad Cooper minimum located just above the photoionization threshold \cite{Fano68}.  This is in sharp contrast with the situation at low $n$. For example, using the measured values of $f_d(n=4)$ \cite{Niemax77} and $f_p(n=5)$ \cite{Shabanova84}, the ratio is $2.25 \times 10^{-6}$, two orders of magnitude smaller than at high $n$.

\section{Ratio of $J$-Dependent Oscillator Strengths}
\begin{figure}
\centering \vskip 0 mm
\includegraphics[width=0.95\linewidth]{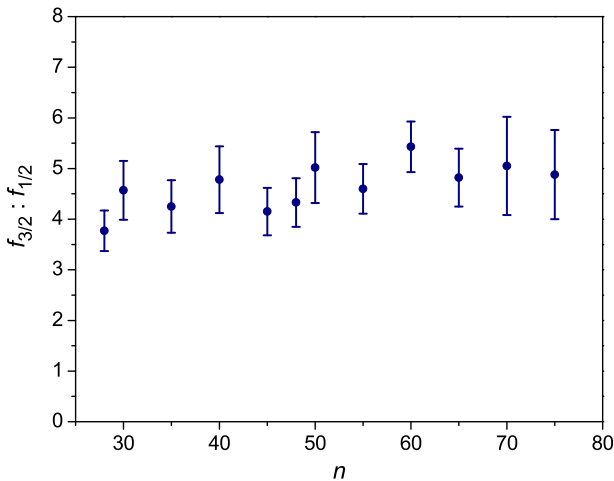}
\caption{\protect\label{fig5} (Color online) Measured n dependence of ratio $\rho = f_{3/2}/f_{1/2}$ of oscillator strengths for transitions $5s_{1/2} \rightarrow np_{3/2}$ and $5s_{1/2} \rightarrow np_{1/2}$.}
\end{figure}

Our narrow-band UV excitation of an ultracold sample allows us to resolve the Rb $np$ fine-structure splitting up to $n$=75. From these spectra, an example of which is shown in Fig. 1, we can obtain the ratio of fine-structure E1 oscillator strengths, $\rho(n) = f_{3/2}/f_{1/2}$, for transitions $5s_{1/2} \rightarrow np_{3/2}$ and $5s_{1/2} \rightarrow np_{1/2}$. This ratio is taken to be the ratio of areas under the respective spectral peaks. The variation of $\rho$ with $n$ is shown in Fig. 5. The uncertainties shown are statistical.  For some $n$ values, only a small number of measurements (e.g., 3) were taken, leading to a large scatter in statistical estimates of the uncertainty.  Therefore, the final uncertainties were assigned by averaging the uncertainties for adjacent values of $n$.  Over the range of $n$ explored, 28 to 75, $\rho$ is rather constant.

A weighted average of this data yields $\rho = 4.50 \pm 0.16$  This is clearly inconsistent with the statistical value of 2.0 based on the 2$J$+1 degeneracies alone. This anomaly has been previously noted, and is attributed to the interplay of the spin-orbit interaction, core polarization, and cancelation effects in transition dipole moment matrix elements \cite{Migdalek98}. There are several previous experimental measurements of the ratio $\rho$ at low $n$ \cite{Shabanova84,von_der_Goltz84,Caliebe79} and one at high $n$ \cite{Liberman79}. It has the expected statistical value of 2.0 at $n$=5 \cite{Shabanova84}, then rises with increasing $n$ before leveling off above $n$=20. The highest $n$ for which an individual value has been previously reported is $\rho(n=20) = 4.9 \pm 0.2$ \cite{Caliebe79}. A value of $\rho(n=25) = 5.1$ is quoted in \cite{von_der_Goltz84}. The average value over the range $n=29-50$ was determined to be $\rho = 5.9 \pm 1.4$ \cite{Liberman79}. All of these results are consistent with our present measurements.

A number of calculations of $\rho$ have been performed at low $n$ ($\leq10$) \cite{Warner68,Migdalek79}, intermediate $n$ ($\leq$25) \cite{Weisheit72,Hofsaess77,Hansen84,Migdalek98}, and high $n$ ($\leq$80) \cite{Luc-Koenig78}. The predictions of \cite{Weisheit72} and \cite{Hofsaess77}, when extrapolated to high $n$, are significantly higher than our measurements. The extrapolated results of \cite{Hansen84} appear to be marginally consistent but slightly higher. The variational and frozen-core calculations of \cite{Luc-Koenig78} agree well with our results. Our measurements are also consistent with the extrapolated RMP+CPIB results of Migdalek \cite{Migdalek98}. This calculation uses a relativistic model potential and incorporates core polarization effects in both the potential and the transition dipole moment operator.

We also measure the ratio $f_{5/2}/f_{3/2}$ of E2 oscillator strengths for the $5s_{1/2} \rightarrow nd_{5/2}$ and $5s_{1/2} \rightarrow nd_{3/2}$ transitions. Because the $nd$ fine-structure splittings are smaller than for $np$, we are only able to resolve the $nd_J$ levels up to $n$=48. We find that $f_{5/2}/f_{3/2} = 1.56 \pm 0.07$, independent of $n$ over the range $n=28-45$. This is consistent with the statistical value of 1.5. To the best of our knowledge, there are no calculations of this E2 oscillator strength ratio, but there is also no reason to expect non-statistical behavior.

\section{Conclusions}
In summary, we have measured oscillator strengths for electric quadrupole transitions to highly-excited $nd$ states of Rb. We find that these E2 transitions are weaker than the E1 transitions to nearby $np$ states by a slightly $n$-dependent factor of only $\sim2000$. We have also determined the relative E1 transition strengths to the two $np$ fine-structure states, $J$=3/2 and $J$=1/2. For the states we have investigated, the measured ratio of $4.50 \pm 0.16$ is independent of $n$ and differs dramatically from the statistical value of 2.0 expected from degeneracies alone. Both anomalies, the unexpectedly large E2 to E1 transition strength ratio and the non-statistical $np$ fine-structure transition strength ratio, owe their origin in part to the pronounced Cooper minimum \cite{Fano68} for $5s \rightarrow np$ transitions that lies just above threshold in Rb \cite{Migdalek98}.

\begin{acknowledgments}
We gratefully acknowledge funding support from NSF awards PHY-0457126 and PHY-0653449.
\end{acknowledgments}

\end{document}